\documentclass[]{spie}  %>>> use for US letter paper
\usepackage[]{graphicx}

\def\arcmin{\hbox{$^\prime$}}
\def\arcsec{\hbox{$^{\prime\prime}$}}
\def\deg{\hbox{$^\circ$}}

\def\lae{\mathrel{\raise .4ex\hbox{\rlap{$<$}\lower 1.2ex\hbox{$\sim$}}}}
\def\gae{\mathrel{\raise .4ex\hbox{\rlap{$>$}\lower 1.2ex\hbox{$\sim$}}}}

\title{A Soft X-ray Polarimeter Designed for Broad-band X-ray Telescopes} 

\author{Herman L. Marshall\supit{a}
\skiplinehalf
\supit{a}MIT Kavli Institute, Cambridge, MA, USA
}

\authorinfo{Further author information: E-mail: hermanm@space.mit.edu, Telephone: 1 617 253 8573} 

\begin{document} 
  \maketitle 

%%%%%%%%%%%%%%%%%%%%%%%%%%%%%%%%%%%%%%%%%%%%%%%%%%%%%%%%%%%%% 
\begin{abstract}
A novel approach for measuring linear X-ray polarization over a broad-band using conventional imaging optics and cameras is described.
A new type of high efficiency grating, called the critical angle transmission grating is used to disperse soft X-rays radially from the telescope axis.  A set of multilayer-coated paraboloids re-image the dispersed X-rays to rings in the focal plane.  The intensity variation around these rings is measured to determine three Stokes parameters: I, Q, and U.  By laterally grading the multilayer optics and matching the dispersion of the gratings, one may take advantage of high multilayer reflectivities and achieve modulation factors over 50\%
over the entire 0.2 to 0.8 keV band.  A sample design is shown that could be used with the
Constellation-X optics.  
\end{abstract}

%>>>> Include a list of keywords after the abstract 


%%%%%%%%%%%%%%%%%%%%%%%%%%%%%%%%%%%%%%%%%%%%%%%%%%%%%%%%%%%%%
\section{INTRODUCTION}
\label{sec:intro}  % \label{} allows reference to this section

Although polarization is a fundamental property of electromagnetic radiation,
there have been no astronomical measurements of polarization in the X-ray
band in over 30 yr.  There has been no lacking for workable concepts, including
simple designs such as the Polarimeter for Low Energy X-ray Astrophysical
Sources (PLEXAS) that was proposed by Marshall et al.\ to use
multilayer-coated mirrors tuned to 0.25 keV as Bragg reflectors.\cite{plexas}
An excellent review of the history and prospects for astronomical
polarimetry in the 0.1-10 keV band is presented by Weisskopf et al.\cite{weisskopf06}.
Weisskopf et al.\ rightly argue that the PLEXAS
design has a narrow bandpass, reducing its attractiveness
for astrophysical observations because one expects polarization to be energy
dependent, so a wide bandpass is desired.

Here, I outline a new polarimeter design that can take advantage of broad-band X-ray optics
by dispersing the incoming X-rays onto a multilayer-coated secondary.  The graze
angles off of the secondary are large, of order 30-40\deg, and the polarization is determined
in aggregate by measuring intensity around rings in the focal plane.  Compared to the
PLEXAS design, the bandpass increases from about 20 eV to over 500 eV.
The basic approach is very flexible.  For the sake of demonstration, a configuration is
designed for the Constellation-X mission.  A similar design could be used for the XEUS mission
or be designed for a small explorer.

\section{Primary Mirror and Gratings}

The approach to this polarimeter design was inspired by a new blazed transmission grating
design, called the Critical Angle Transmission (CAT) grating \cite{heilmannspie07}, and the
corresponding application to the Constellation-X (Con-X) mission in the design of
a transmission grating spectrometer\cite{flanagan07}.
For details of the Con-X mission, see the Con-X web site at
{\tt http://constellation.gsfc.nasa.gov/}.  This type
of grating gives very high efficiency in first and higher orders.  With such high effiiciency,
a dispersed spectrum could be matched to a multilayer-coated secondary mirror that
would then perform the function of a polarizer.

We assume that the primary mirror system is capable of focussing X-rays over
the 0.1 to 1.0 keV band.
For Con-X, the baseline optics are four assemblies consisting of
$n$ conical shells with
an outer radius $r_1 = 0.65$ m and an inner radius
$r_n = 0.15$ m.  The effective area goal is 3.8 m$^2$ using 4 independent
mirror assemblies.  The focal length $F = 10.0$ m.
The mirror resolution goal is $\delta \theta = 5$\arcsec, while the baseline is 15\arcsec.

\section{Multi-Detector Approach}

\begin{figure}
   \begin{center}
   \begin{tabular}{c}
   \includegraphics[height=17cm]{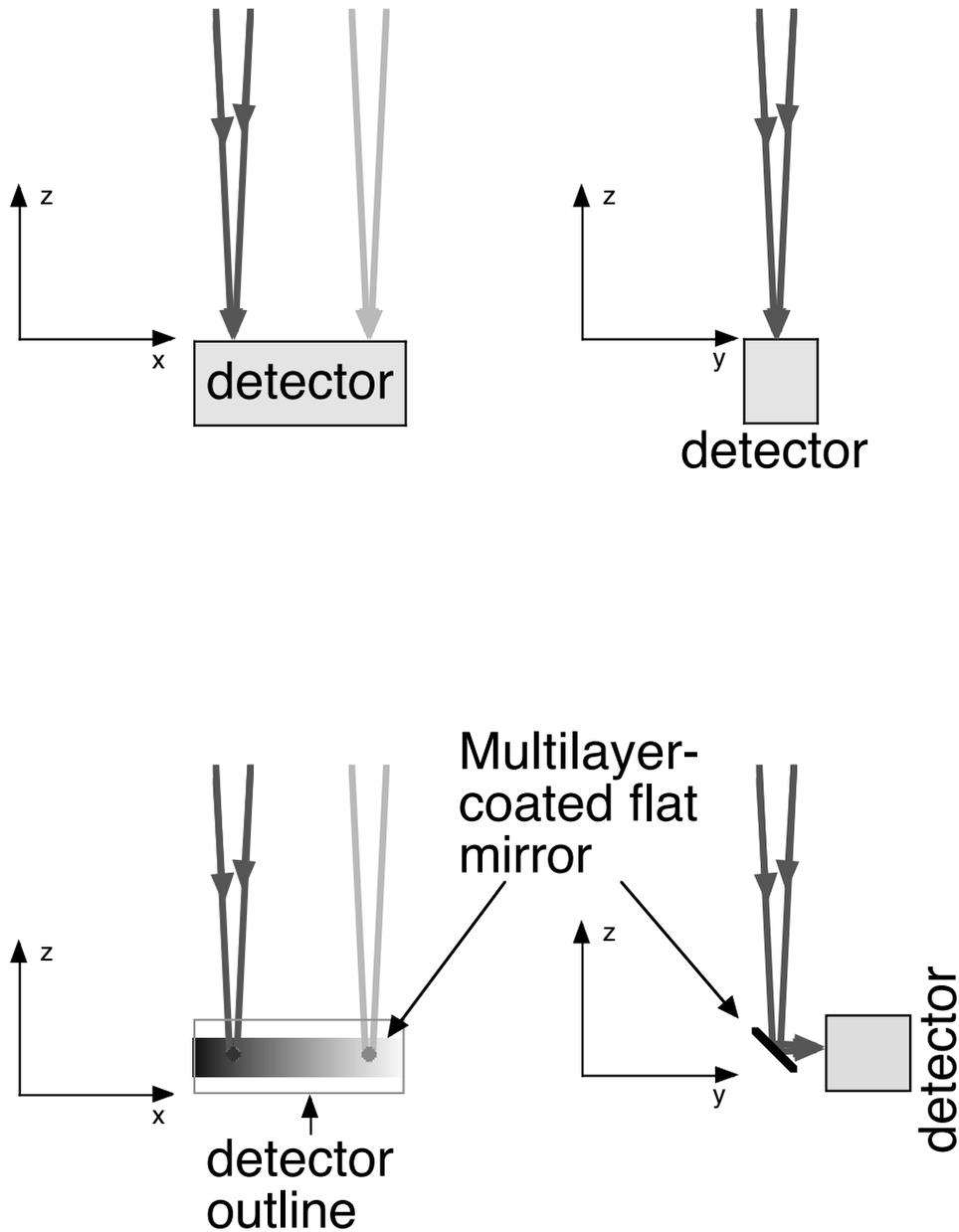}
   \end{tabular}
   \end{center}
   \caption
   { \label{fig:conxlayout} 
 {\it Top:} Schematic of a single detector receiving dispersed X-rays from gratings behind the
 telescope, as in Constellation-X.  Two views are shown, where the $z$ axis is along the telescope
 axis.  The spectrometer dispersion axis is approximately parallel to the $x$ axis, along the surface
 of the detector (which is segmented to follow the Rowland circle).
 The cross-dispersion direction is parallel to the $y$ axis.  Two different wavelengths
 of X-rays arrive at the focal plane, differing in $x$ according to the grating dispersion relation.
 Only one of these is shown in the right-hand view.  {\it Bottom:} Same telescope and grating optics
 but now a polarizing flat has been added.  The flat is placed at the Brewster angle (45\deg) to the incoming
 X-rays to null the reflectivity of $p$-polarized X-rays.  The multilayer coating
 period varies with position along the $x$ axis according to
 Eq.~(\ref{eq:dspace}), tuned to reflect short
 wavelengths at low $x$ and long wavelengths at large $x$.  The detector is rotated
 90\deg\ about the $+x$ axis so that the focal plane is now approximately parallel to the $x-z$
 plane.  In the left-hand view, the detector would block the view of the multilayer-coated flat
 mirror, so only an outline of the detector is shown.  Note that the flat and the detector are
 displaced in the $+z$ direction from the original focal plane.}
   \end{figure}

One could convert the CAT-based spectrometer to a polarimeter very
simply by adding multilayer-coated flats at the current locations of the readout
CCD arrays to reflect the X-rays by 90\deg.  The detectors would be repositioned
and rotated 90\deg\ to receive the reflected X-rays, facing these flats.
Fig.~\ref{fig:conxlayout} shows how the focal plane might look.
The multilayer spacing, $d$, would
vary laterally (along the dispersion) in order to provide optimal reflection
at a graze angle $\theta = $45\deg.
Such optimization is obtained by equating the wavelength from
the linearized grating equation, $m \lambda = P x/D_R$, to that of a multilayer,
$\lambda = 2 d \sin \theta$, giving

\begin{equation}
\label{eq:dspace}
d = \frac{Px}{2^{1/2} m D_R} \,  ,
\end{equation}
where $x$ is the distance along the dispersion
from 0th order, $D_R$ is the Rowland diameter
(the distance between the gratings and the detector plane), $m$ is the
grating order, and $P$ is the period of the gratings.
Thus, the multilayer period varies linearly with $x$, providing high reflectivity
in a narrow bandpass at large graze angles.  For a given order,
the spectrometer's gratings direct a narrow bandpass to any given position
on the flat, so the bandpass of the multilayer can be similarly narrow.
The bandpass of the multilayer decreases approximately as $1/N$ for a large
number of layers, $N$, so if the spectrometer has a resolution better than 100,
say, then $N \gae 100$.  Large $N$ also improves the multilayer reflectivity.

At Brewster's angle, $\theta = 45$\deg, reflectivity is minimized when the $E$-vector
is in the plane containing the incident ray and the surface normal ($p$-polarized)
and maximized when the $E$-vector is in the surface plane ($s$-polarized).
The polarization position angle (PA) is the average orientation in sky coordinates
of the $E$-vector for the incoming X-rays.
Sampling at least 3 PAs is required in order to measure
three Stokes parameters (I, Q, U) uniquely, so one would require at
least three separate detector
systems with accompanying multilayer-coated flats.  This design is more costly and heavy than
the original design by Flanagan et al.\cite{flanagan07} due to the need
for a third detector system.  Another problem is that $m$ has to be chosen in order
to set the multilayer spacing, reducing efficiency in high orders, which are critical
to achieving high throughput at short wavelengths.\cite{flanagan07}
For these reasons, the rest of this paper will focus on a design that requires
only one detector to measure the polarized signal and sets $m = 1$.

\section{Single Detector System}

\begin{figure}
   \begin{center}
   \begin{tabular}{c}
   \includegraphics[height=15cm,angle=270]{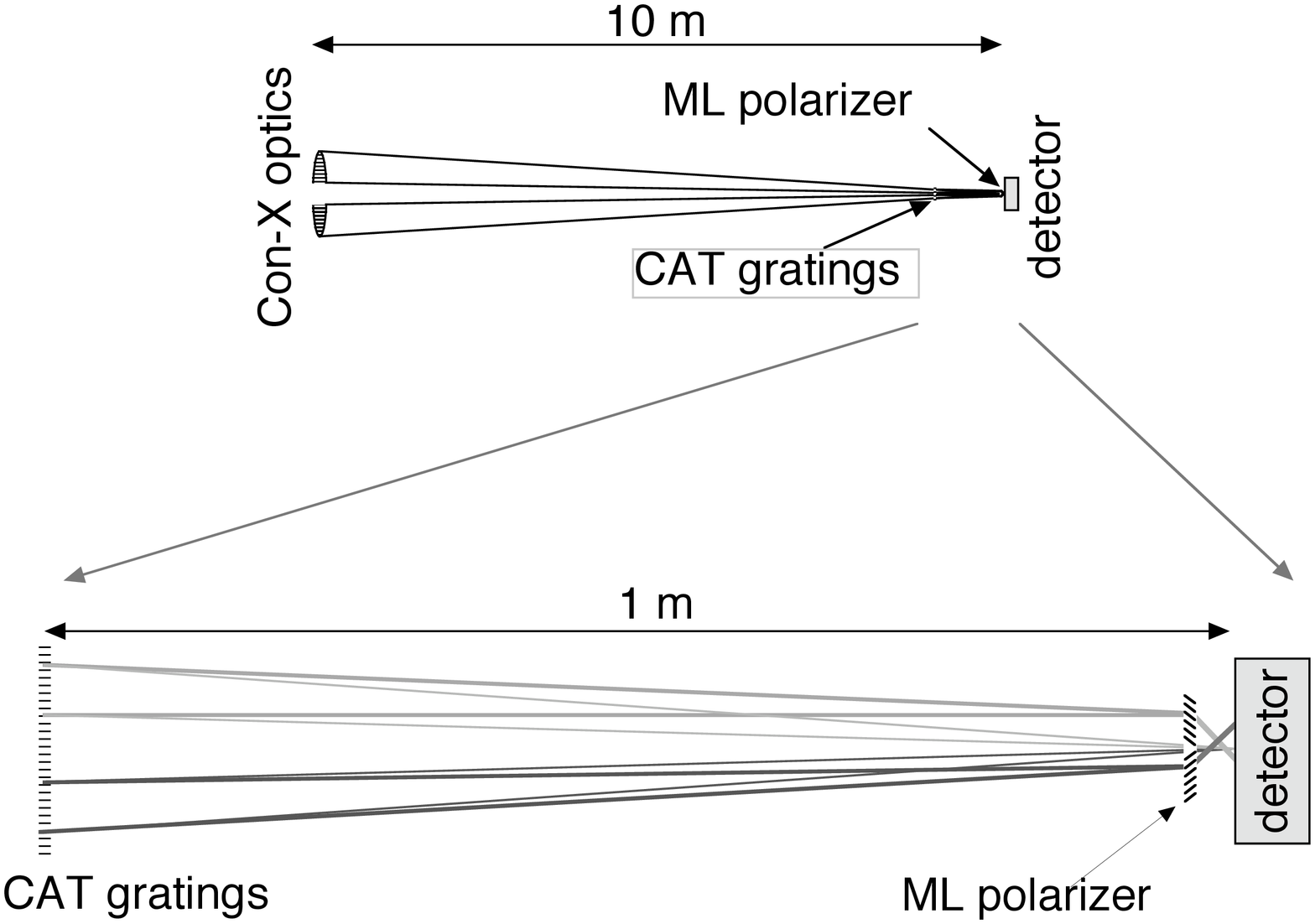}
   \end{tabular}
   \end{center}
   \caption
   { \label{fig:layout} 
 {\it Top:} The Constellation-X optical system with Critical Angle Transmission
 (CAT) gratings\cite{heilmannspie07} and
 multilayer (ML) polarizing optics added,
approximately to scale (except for the detector size).
The CAT gratings are 1 m from the focal plane
and the multilayer polarizers are 15-60 mm from the detector.  {\it Bottom:}
Blowup of the polarizing optics.  The CAT gratings disperse away from the telescope focus;
0th order light (thin lines) continue to the focus. 
Low energy (light gray) or high energy (dark gray) X-rays are dispersed so that wavelength
increases from the optical axis.
The polarizer reflects at large graze angles so that the intensity around the ring provides
polarization information.}
   \end{figure}

If there is to be only one detector that is oriented perpendicular to
the telescope axis, then the gratings can be designed to disperse the focussed X-rays away from the
telescope axis and a set of multilayer-coated mirrors would redirect the light to the
focal plane.  This design is illustrated schematically in Fig.~\ref{fig:layout}.  In this
case, the gratings are arranged to disperse radially from the optical axis, so that
wavelength increases approximately linearly with distance $x$ from the optical axis.

%% tabular environment useful for creating an array of images  
   
\subsection{Disperser}

\begin{figure}
   \begin{center}
   \begin{tabular}{c}
   \includegraphics[height=11cm]{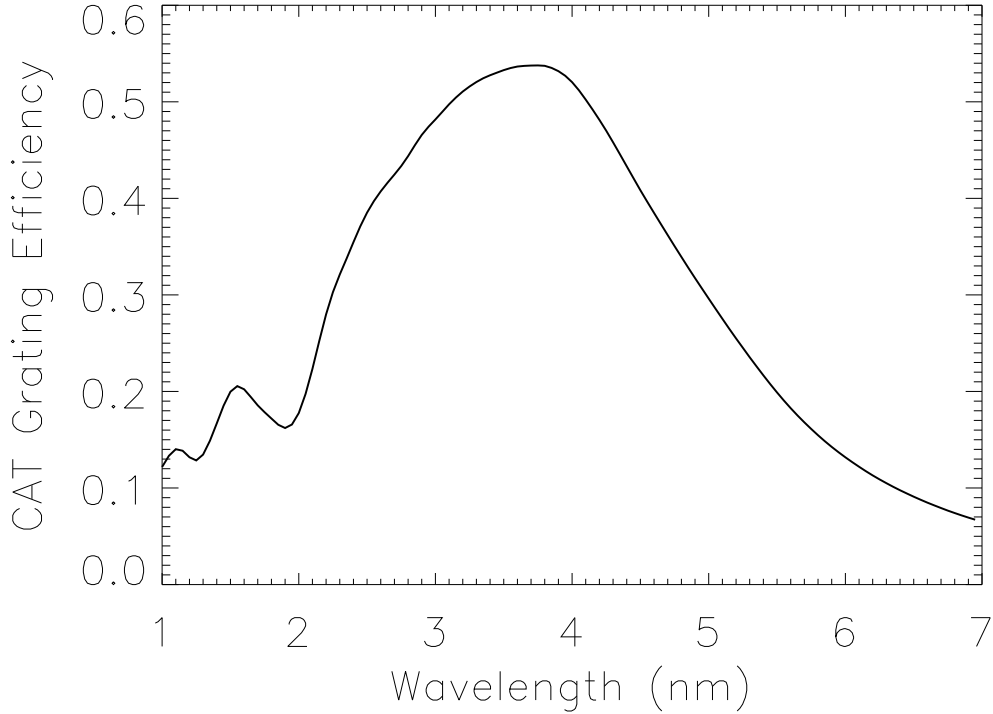}
   \end{tabular}
   \end{center}
   \caption
%>>>> use \label inside caption to get Fig. number with \ref{}
   { \label{fig:cateff} 
Computed efficiency of the CAT gratings\cite{heilmannspie07} in first order used in this study
(R. Heilmann, private communication).
The CAT grating design has $P = 100$ nm, bars that are 3.5 $\mu$
tall, 15 nm wide, and tilted at 0.95\deg\ to the vertical.  The efficiency peaks
over 50\%.}
   \end{figure}

Transmission gratings are used to disperse the X-rays.  The main
requirements are to obtain a spectral resolution $R \equiv E/\delta E \gae N \sim 100$ and to
have high efficiency.  The medium
energy gratings (MEGs) of the {\it Chandra} X-ray Observatory
High Energy Transmission Grating Spectrometer\cite{hetgs} could be used, for example.
The MEGs have a polyimide supporting membrane that reduces throughput between
0.3 and 0.5 keV, so a grating like that used for the {\it Chandra} Low
Energy Transmission Grating Spectrometer\cite{letgs} (LETGS) might be
more appropriate.
However, the peak efficiency in $+1$ order is about 10\% for these gratings,
while that of the blazed CAT gratings can reach 50\%\cite{heilmannspie07}.
The CAT grating design used in this study has $P = 100$ nm, bars that are 3.5 $\mu$
tall, 15 nm wide, and tilted at 0.95\deg\ to the vertical.  The predicted
first order efficiency (R. Heilmann, private communication),
shown in Fig.~\ref{fig:cateff}, peaks at over 50\%.
If placed just behind the Con-X mirrors, the CAT grating structure would have to
be about 1.3 m across and the resolution at the focal plane would be $\gae$ 1000.
This polarimeter design only requires $R \sim 100$, however, so the grating
assembly can be placed about 1 m from the focal plane, reducing the size to
about 130 mm diameter.  The mass of the assembly is reduced by a factor of 100.
For this period and assuming $D_R = 1$ m, the
grating equation gives $x = \lambda$ for $x$ in mm
and $\lambda$ in \AA.

\subsection{Multilayer Polarizer}

\begin{figure}
   \begin{center}
   \begin{tabular}{c}
   \includegraphics[height=15cm,angle=90]{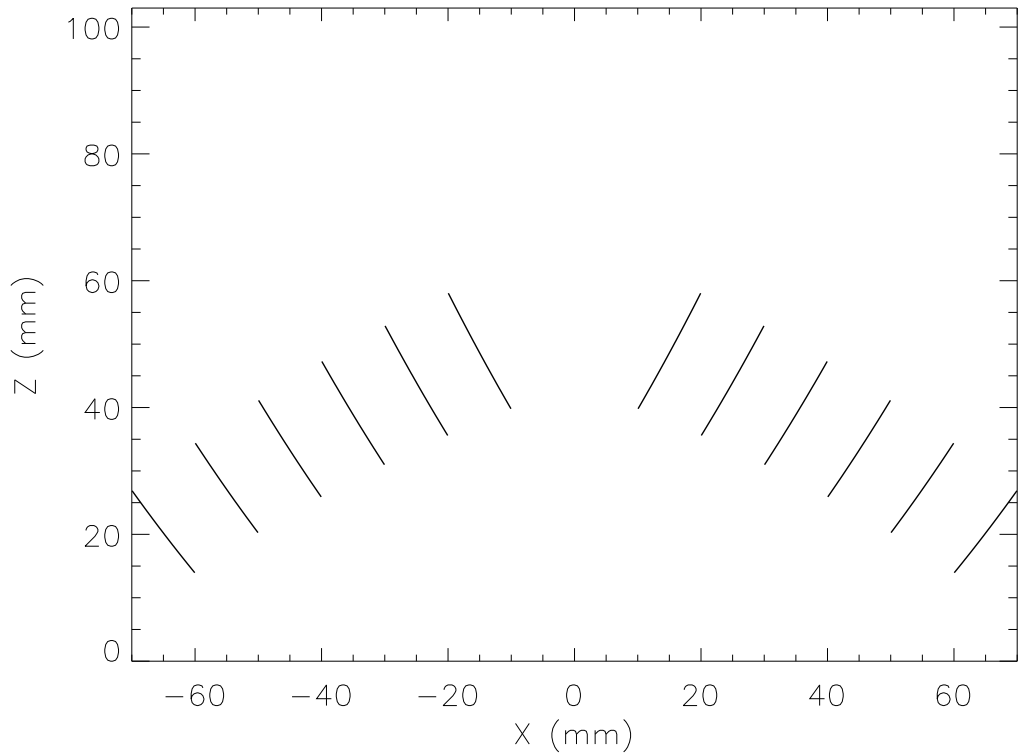}
   \end{tabular}
   \end{center}
   \caption
%>>>> use \label inside caption to get Fig. number with \ref{}
   { \label{fig:mlring} 
Cross section of the nested paraboloids that are multilayer coated to
reflect with high efficiency in specific wavelength bands.
The paraboloids redirect X-rays from the gratings
into narrow rings in the focal plane (see Fig.~\ref{fig:focplane}). 
Each surface has a laterally graded multilayer that is tuned to
the wavelength arriving there from the grating.
Other configurations were studied and are not shown.  The innermost
ring intercepts X-rays with wavelengths between 1 and 2 nm, the next
ring reflects 2-3 nm X-rays, etc.}
   \end{figure}

The final optical element is a set of nested reflectors that are approximately
paraboloidal.  A cross section through the rings is shown in Fig.~\ref{fig:mlring}.
The ring spacing was chosen to reflect the radially dispersed X-rays into
a ring image.  Each paraboloid has a different image ring size and reflects
a different 1 nm bandpass in this design.  Highest efficiency
is achieved when a narrow bandpass from the gratings matches
the narrow reflectivity response of the multilayer.  The reflection angles
of the paraboloids are set in order to obtain graze angles between 25
and 40\deg\, from considerations of the modulation factor
and from the requirement that any given paraboloid
does not occult the light reflected off of another.

\begin{figure}
   \begin{center}
   \begin{tabular}{c}
   \includegraphics[height=10cm]{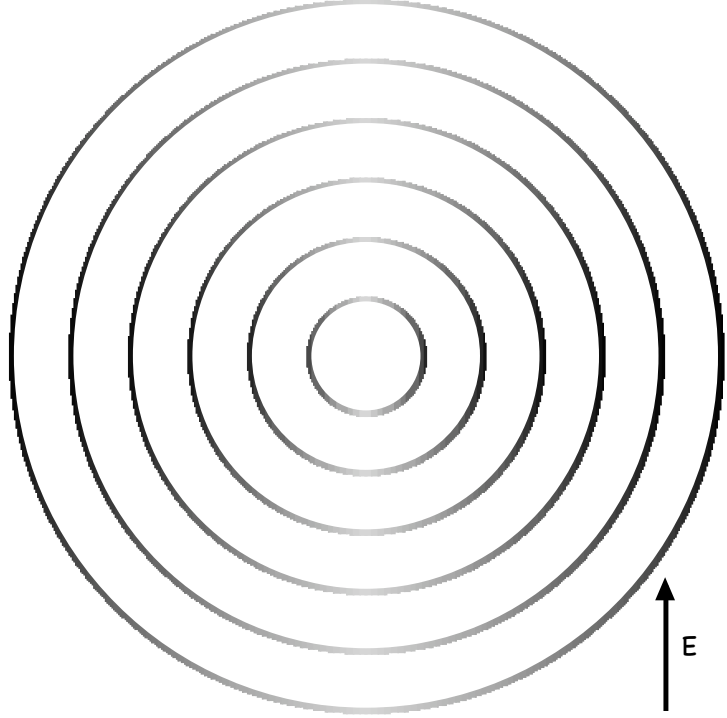}
   \end{tabular}
   \end{center}
   \caption
   { \label{fig:focplane} 
 Ring images produced by the multilayer optics. Each ring corresponds to a different
 wavelength interval.  The widths of the ring images are related to the mirror
 angular resolution and are generally $<10$\% of the distance between rings.
 The intensity around each ring varies as $\cos (2 [\phi - \varphi] )$, where $\phi$
 is the azimuthal angle, and $\varphi$ determined by the $E$-vector of
 the polarized X-rays.  Here, $E$ is parallel to the $y$ axis for each
 ring, so ring intensity is
 maximized along the $x$ axis.  In general, the minima of each ring may differ,
 indicating a wavelength-dependent polarization position angle.
 The relative intensities of the rings depends
 on the source spectrum and the system throughput; here,
 a flat spectrum is assumed.}
   \end{figure}

The images that
might be produced are shown schematically in Fig.~\ref{fig:focplane}.
In this figure, the outermost ring image is generated by the
innermost paraboloid: the 1-2 nm wavelength range (0.62 to 1.24 keV).  Conversely,
the innermost ring image is formed by the most highly dispersed X-rays,
with wavelengths between 6 and 7 nm (0.18 to 0.21 keV).
The nested reflectors should  be coated so as to reflect only those X-rays
at the appropriate Bragg condition.  The graze angles are shown in
Fig.~\ref{fig:dspace}, along with the required multilayer period, $d$.
Following Marshall et al.\cite{pkspolar},
the intensity $I_i$ around ring image $i$ in the $\delta \lambda$ band about
$\lambda_{i}$ varies with azimuthal angle $\phi$ as

\begin{equation}
I_i\left({\phi }\right)=\ {I}_i^{0}\left\{{{\frac{1}{2\pi }}\ +\ {P_i^0} M
	\cos \left({2\left[{\phi \rm -\varphi_i }\right]}\right)}\right\}\ \ ,
\end{equation}

\noindent
where $I_i^0$ is directly related to the I Stokes parameter of the source
for bandpass $i$,
$P_i^0$ is the average polarization fraction across the wavelength
band, $\varphi_i$ is the phase determined by the orientation
of average polarization $E$-vector on the sky (i.e., the polarization PA),
and $M_i$ is the average system modulation factor in bandpass $i$.
At any given wavelength, $M = \frac{R_s-R_p}{R_s+R_p}$,
where $R_s$ and $R_p$ are the
reflectivities for the $s$ and $p$ polarizations, respectively.
The Stokes Q and U parameters can be derived from $P_0$ and
$\varphi$ as a function of wavelength.

\subsection{System Design Issues}

\begin{figure}[t]
   \begin{center}
   \begin{tabular}{c}
   \includegraphics[height=11cm]{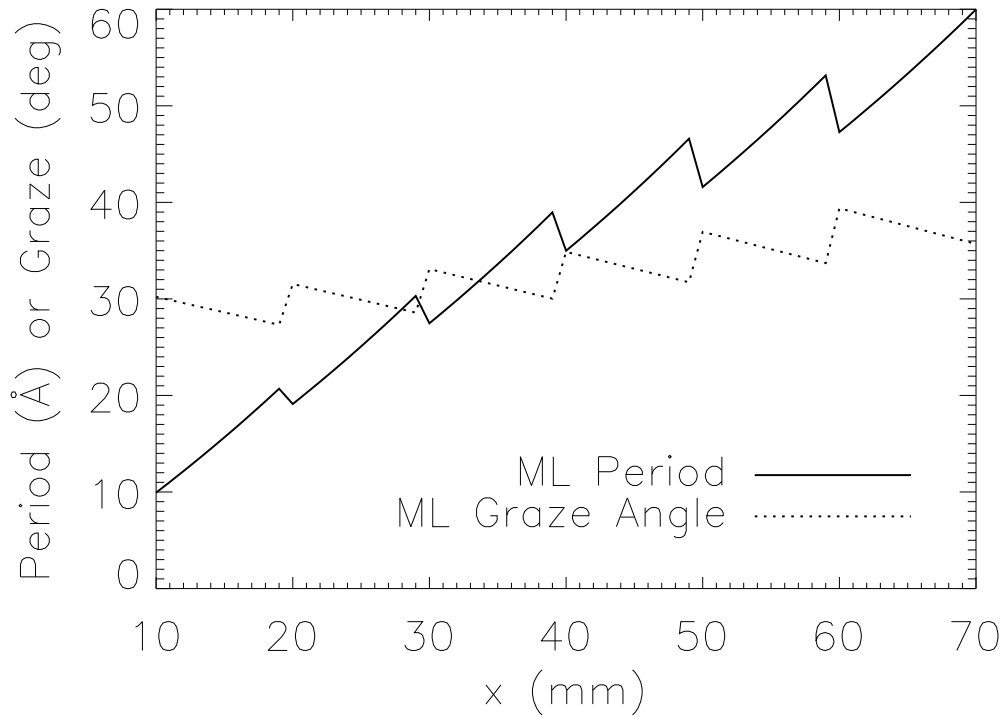}
   \end{tabular}
   \end{center}
   \caption
%>>>> use \label inside caption to get Fig. number with \ref{}
   { \label{fig:dspace} 
Variation of the multilayer period, ($d$, solid line), and the
graze angle ($\theta$, dotted line) with X-ray wavelength.  For the assumed
grating period and placement of the grating from the focal plane,
X-rays with wavelength $\lambda$ in \AA\ disperse to distance
$x = \lambda$, for $x$ in mm.  The discontinuities result from
segmenting the multilayer-coated polarizer.}
   \end{figure}

The widths
of the rings would be limited by the optical system, including the
multilayer-coated optics.  For Con-X with 5\arcsec\ primary optics, the rings would be at least
0.25 mm wide, so there is some room in the focal plane between ring images
for background measurement and additional sources in the field.
The field of view would be limited by the condition that the change of
the graze angle on the paraboloids keep the X-ray within the spectral
response band of the $N$-layer coating,
$R  \sim N \sim 100$.
Differentiating the condition on the multilayer central wavelength, we find
$2 d  \cos \theta ~ \delta \theta = \delta \lambda$, giving $\delta \theta =
\tan \theta \delta \lambda / \lambda = \tan \theta / R \sim 0.01 \tan \theta$.  For $\theta$ given
in Fig.~\ref{fig:dspace}, $\delta \theta = 0.3-0.5$\deg\ -- much larger than
the distance between the rings, which corresponds to 3.5\arcmin.

More important is the spread of the focussed X-rays as they exit
the primary mirror assembly.  In this design, where only one order
is used, the polarizer receives X-rays over an angular range
$\frac{r_1 - r_n}{F} = 0.05$ rad,
or about 2.9\deg.  The spread should be narrower
than $\frac{\tan \theta}{N}$, so the innermost useful
radius at the telescope aperture cannot be as small
as 0.15 m as in the Con-X baseline.  The gratings need not
intercept the full incoming beam, however.  The innermost
radius corresponds to $r_1 - \frac{F \tan \theta}{N} = 0.51$ m
for an average graze angle of 35\deg.  Limiting the beam this
way cuts the effective area by a factor of 2.5.  Alternatively,
if the number of multilayers was reduced to $N = 70$, say, there
would be about a 35\% gain in usable aperture, offset by
a $\sim$ 30\% loss of peak reflectivity.  The multilayer reflectivity
losses depend on wavelength, so a tradeoff study would be required.

The inner region of the focal plane is not used for this polarimeter
design.  This region is about 3.5\arcmin\ in diameter,
which satisfies the Con-X field of view requirement:
5\arcmin\ $\times$ 5\arcmin.
This design could be used in conjunction with a high
energy non-dispersive spectrometer because the gratings
have a high 0th order efficiency, approaching 100\% above
2 keV.  Thus, low energy polarimetry would not interfere
with high resolution spectroscopy, the primary objective of the
Con-X mission, and the data would be obtained
simultaneously.

\subsection{System Throughput and Sensitivity}

Using the Center for X-ray Optics web page
({\tt http://henke.lbl.gov/optical\_constants/multi2.html}),
multilayer efficiencies were computed for a small range of materials for
many of the relevant graze angles and corresponding $d$ values.
For this work, $N$ was set to 100 and the interdiffusion
thickness was set to 0.3 nm, as has been achieved by several
groups\cite{windt04,1998ApOpt..37.1873M,1996ApOpt..35.5134M}.
Table~\ref{tab:mlspecs} gives a sampling of the results, where
$r$ is the peak reflectivity to unpolarized X-rays, and $\Gamma$
gives the fraction $d$ that is occupied by the first material given.

\begin{table}[h]
\caption{Sample Multilayer Specifications} 
\label{tab:mlspecs}
\begin{center}       
\begin{tabular}{|c|ccccccc|} %% this creates two columns
\hline
$\lambda$	&	$E$	&	$\theta$	&	$d$	&	$r$	&	$M$	&	Materials	&	$\Gamma$	\\	
(nm)	&	(keV)	&	(\deg)	&	(nm)	&		&		&		&		\\	
\hline
1.3	&	0.954	&	29.2	&	1.34	&	0.03	&	0.56	&	NaBr/Ir	&	0.6	\\	
1.7	&	0.729	&	28.0	&	1.82	&	0.08	&	0.49	&	NaF/W	&	0.6	\\	
2.0	&	0.620	&	31.5	&	1.92	&	0.10	&	0.62	&	NaF/Ni	&	0.6	\\	
2.5	&	0.496	&	29.9	&	2.53	&	0.14	&	0.53	&	SiO$_2$/Cr	&	0.6	\\	
3.2	&	0.387	&	32.4	&	2.99	&	0.27	&	0.53	&	Sc/Cr	&	0.6	\\	
3.8	&	0.326	&	30.4	&	3.81	&	0.19	&	0.48	&	Sc/Cr	&	0.6	\\	
4.0	&	0.310	&	34.9	&	3.54	&	0.15	&	0.67	&	Sc/Cr	&	0.6	\\	
4.4	&	0.282	&	33.5	&	4.05	&	0.14	&	0.53	&	Sc/Cr	&	0.6	\\	
4.5	&	0.276	&	33.1	&	4.16	&	0.29	&	0.56	&	C/Cr	&	0.7	\\	
4.9	&	0.253	&	31.7	&	4.66	&	0.26	&	0.54	&	C/Cr	&	0.7	\\	
5.0	&	0.248	&	36.9	&	4.16	&	0.20	&	0.78	&	C/Cr	&	0.7	\\	
6.0	&	0.207	&	39.4	&	4.83	&	0.16	&	0.88	&	C/Cr	&	0.7	\\	
7.0	&	0.177	&	35.7	&	6.00	&	0.15	&	0.74	&	C/Cr	&	0.7	\\	
\hline 
\end{tabular}
\end{center}
\end{table} 

\begin{figure}[ht]
   \begin{center}
   \begin{tabular}{c}
   \includegraphics[height=15cm]{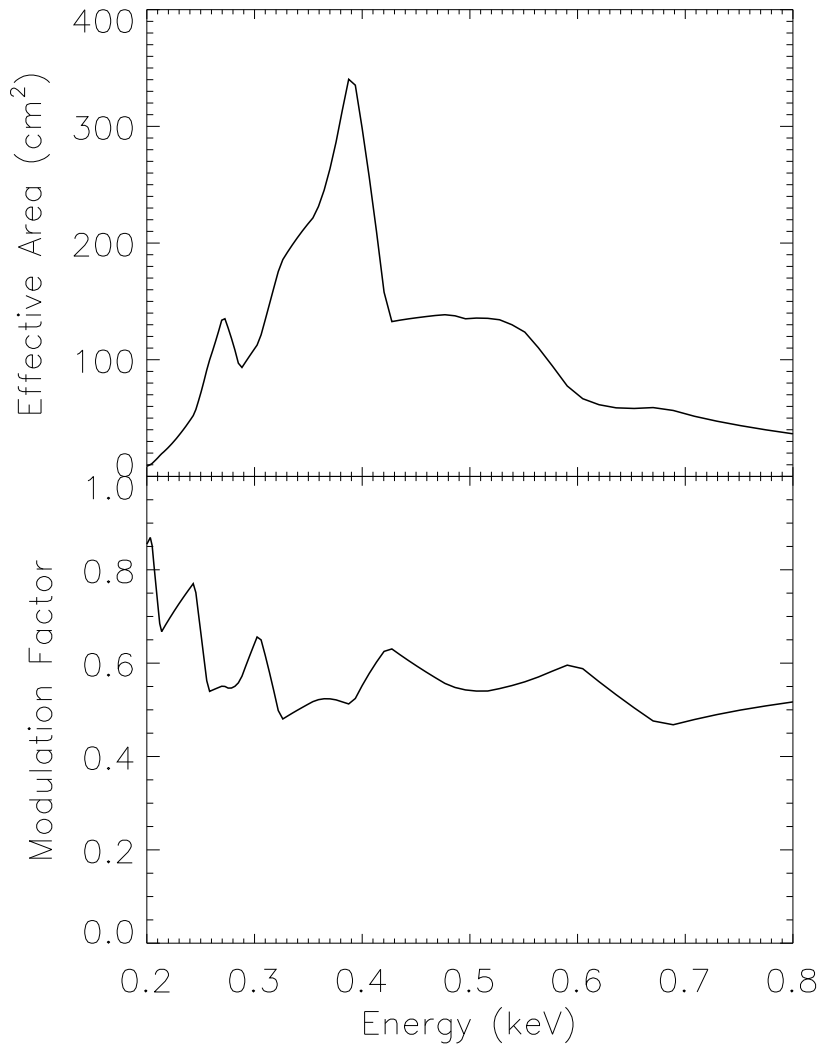}
   \end{tabular}
   \end{center}
   \caption
%>>>> use \label inside caption to get Fig. number with \ref{}
   { \label{fig:effarea} 
Effective area (top) and the polarization modulation factor (bottom) for a system
where one Con-X mirror is used with a CCD detector and a thin optical blocking
filter.  An effective area of over 100 cm$^2$ can be achieved over most of the
0.2-0.8 keV bandpass.}
   \end{figure}

Assuming that this system is used in conjunction with one Con-X telescope
(out of four),
the geometric area in the .2-1 keV band is about 3300 cm$^2$, after accounting
for the factor of 2.5 loss described in the last section.
For the purposes of this analysis, a CCD detector is assumed to have a thin
Al overlayer to act as an optical blocking filter.
The resulting effective area is shown in
Fig.~\ref{fig:effarea}, along with the modulation factor, $M$.

The effective area estimate can be used to predict the minimum
detectable polarization\cite{weisskopf06} (MDP)
for potential targets.  As described by Marshall et al\cite{plexas},
extragalactic sources such as the BL Lac object
PKS~2155$-$304 are expected to be highly polarized in the X-ray band.
In a 10,000 s observation of PKS~2155$-$304,
this instrument could detect polarizations of 3-7\% in
each of four bandpasses 1 nm wide from 2 to 6 nm (0.21 to 0.62 keV).
In addition, isolated neutron stars such as RX J0720.4$-$3125 are expected to polarized
due to effects of photon propagation in strong magnetic fields\cite{weisskopf06}.  In
this case, it is worthwhile to obtain phase-resolved polarization data.  In a 200,000 s
observation with this instrument, one could reach MDPs of 5-8\% over the 2-5 nm
band (in 1 nm wide bands) for each of 10 phase bins.

\subsection{Design Variations}

There is substantial flexibility to this design.  For example, one may sacrifice imaging
if, say, only bright point sources in uncrowded regions were to be observed.  In this case,
the polarizer needn't focus the dispersed X-rays but merely needs to redirect the X-rays
to the detector.  The polarizer could then consist of conic sections (frustums), making them
easier to manufacture and coat.  If the multilayer coating requirements are relaxed,
then one may be able to use unblazed gratings (such as used in the LETGS) and
capture both $+1$ and $-1$ orders, doubling the system throughput at what might be
a small loss in multilayer efficiency.

The grating period need not be as small as 100 nm.  Larger periods may be used but the
structure would have to be correspondingly closer to the mirror, increasing the size and
mass of the assembly.  For example, gratings with $P = 1000$ nm could be placed right
behind the Con-X optics at about 10 m from the focal plane.  For reference, the LETGS
gratings have $P = 991$ nm.\cite{letgs}
The efficiency of the system
is poor above 0.8 keV, where efficient multilayer coatings have not yet been
demonstrated.  Thus, the mirror need not have substantial effective area above 1 keV,
or one may place the gratings only on the telescope perimeter.

There are many combinations of elements and multilayer coating parameters that were
not investigated for this preliminary study, so one might expect to improve upon the
mulilayer-coating reflectivities.  The number of multilayers was held fixed at 100 and no
in-depth $d$ variations were considered, in the interest of simplicity.  It may be possible
to superpolish the substrates to obtain 0.1 nm roughness\cite{2001SPIE.4506..113K},
which would increase reflectivities significantly.

\acknowledgments     %>>>> equivalent to \section*{ACKNOWLEDGMENTS}       
 
This work was supported under contract.  I thank Dr.\ Ralf Heilmann for calculations of
the CAT grating efficiencies for this analysis.

%%%%%%%%%%%%%%%%%%%%%%%%%%%%%%%%%%%%%%%%%%%%%%%%%%%%%%%%%%%%%
%%%%% References %%%%%

\bibliography{polarimeter}   %>>>> bibliography data in polarimeter.bib
\bibliographystyle{spiebib}   %>>>> makes bibtex use spiebib.bst

\end{document}